\documentclass[10pt,cqg]{iopart}

\usepackage{mathrsfs}
\usepackage{iopams}
\usepackage{cite}

\newcommand{\reff}[1]{(\ref{#1})}
\def\uudot{\dot{u}}
\def\3nab{\tilde{\nabla}}

\def\be {\begin{equation}}
\def\ee {\end{equation}}
\def\ba {\begin{eqnarray}}
\def\ea {\end{eqnarray}}
\newcommand{\bra}[1]{\left(#1\right)}
\newcommand{\bras}[1]{\left[#1\right]}
\newcommand{\brac}[1]{\left\{#1\right\}}

\newcommand{\sfr}[2]{{\textstyle\frac{#1}{#2}}}
\newcommand{\fr}[2]{{\frac{#1}{#2}}}
\def\udot{{\cal A}}
\newcommand{\lc}{\varepsilon}
\newcommand{\lb}{\{}
\newcommand{\rb}{\}}
\newcommand{\E}{{\cal E}}

\newcommand{\barray}{\begin{array}}
\newcommand{\earray}{\end{array}}
\newcommand{\n}{n}
\newcommand{\N}{N}
\newcommand{\sdel}{{\mathrm{D}}}

\newcommand{\hatn}{a}
\newcommand{\dotn}{\alpha}
\newcommand{\Si}{{\mathsf{S}}}
\newcommand{\Vi}{{\mathsf{V}}}
\newcommand{\Ti}{{\mathsf{T}}}
\newcommand{\Es}{\mathscr{E}}
\newcommand{\Bs}{\mathscr{B}}
\newcommand{\Js}{\mathscr{J}}

\begin{document}

\title[Scalar field and electromagnetic perturbations on LRS spacetimes]{Scalar field and electromagnetic
perturbations on Locally Rotationally Symmetric spacetimes }

\author{Gerold Betschart\S\dag\ and Chris A. Clarkson\dag\ddag}

\address{\dag\ Department of Mathematics and Applied Mathematics,
  University of Cape Town, 7701 Rondebosch, South Africa}
\address{\S\ Department of Electromagnetics, Chalmers University of
    Technology, SE--412 96 G\"oteborg, Sweden}
\address{\ddag\
  Institute of Cosmology and Gravitation, University of Portsmouth, Portsmouth PO1 2EG, Britain}
\ead{geroldb@maths.uct.ac.za,   chris.clarkson@port.ac.uk}

\date{\today}

\begin{abstract}
We study scalar field and electromagnetic perturbations on Locally
Rotationally Symmetric (LRS) class II spacetimes, exploiting a
recently developed covariant and gauge-invariant perturbation
formalism. From the Klein-Gordon equation and Maxwell's equations,
respectively, we derive covariant and gauge-invariant wave
equations for the perturbation variables and thereby find the
generalised Regge-Wheeler equations for these LRS class II
spacetime perturbations. As illustrative examples, the results are
discussed in detail for the Schwarzschild and Vaidya spacetime,
and briefly for some classes of dust Universes.
\end{abstract}

\pacs{04.20.--q, 04.40.--b}

\section{Introduction}

The covariant 1+3 approach (see \cite{Ellis-vanElst} and
references therein) has proven to be a powerful tool in
relativistic cosmology, especially through its application of the
gauge-invariant, covariant perturbation formalism. This
perturbation formalism works extremely well in cosmological
applications when the background model is homogeneous and
isotropic, that is of Friedman-Lema\^{i}tre-Robertson-Walker
(FLRW) type. However, if the space-time under consideration has
less symmetry, the 1+3 approach is no longer ideally adopted
because its splitting in `time' and `space' relative to the
fundamental observer is not sensitive to another preferred
direction apart from `time'. The description of space-time through
covariant quantities, defined in the observer's restspace, is
simply blind to a second distinguished direction. The 1+1+2
approach \cite{Clarkson-Barrett} remedies this by slicing the
`space' further into `sheets' orthogonal to a second preferred
direction. Analogously, the quantities of the restspace are
further covariantly split in such a way that obtained quantities
still have a clear geometrical or physical meaning. The 1+1+2
approach thus naturally extends the 1+3 approach and keeps its
benefits.

In this article, we illustrate the 1+1+2 formalism with its
application to scalar and electromagnetic perturbations on
spacetimes which are \emph{locally rotationally symmetric} (LRS)
\cite{Ellis67,Stewart-Ellis} possessing a continuous isotropy
group at each point and hence a multiply-transitive isometry
group. Since LRS spacetimes exhibit locally a preferred spatial
direction, the 1+1+2 formalism is ideally suited for a covariant
description of these spacetimes, yielding a complete
characterisation in terms of invariant scalar quantities which
have either a clear physical or direct geometrical meaning. Such a
covariant classification of LRS perfect fluid spacetime geometries
has already been presented in \cite{vanElst-Ellis}, whereas
orthonormal frame methods have been employed in
\cite{Marklund,Marklund-Bradley,Mustapha-Ellis-vanElst-Marklund}.
We will include LRS imperfect fluids in our treatment but the
emphasis will be mainly on LRS class II spacetimes, that is LRS
spacetimes without vorticity terms such that the sheet becomes a
genuine 2-surface.

Schwarzschild black hole perturbations are well understood and it
has been known for a long time that they are all governed by
master equations known as the Regge-Wheeler equation \cite{RW}, a
Schr\"odinger equation with a slightly different potential for
scalar, electromagnetic and gravitational perturbations,
respectively~\cite{RW,PriceI,PriceII,Chandra,Nollert,Kokkotas-Schmidt},
due to the differing spins of the perturbing fields. Using the
1+1+2 formalism, we find the covariant generalisation of the
Regge-Wheeler equation for scalar perturbations, as described by
the Klein-Gordon equation, for all LRS spacetimes, and present the
generalised Regge-Wheeler equation for electromagnetic
perturbations, governed by Maxwell's equations, in the case of LRS
class II spacetimes. We discuss the resulting wave equations in
detail for Schwarzschild and Vaidya
spacetimes~\cite{Kramer-Stephani-MacCallum-Herlt,Vaidya1}, the
latter being closely related to the former by having a
non-increasing mass. We also describe (source-free)
electromagnetic perturbations on the Schwarzschild geometry by a
linear first order system of ODEs plus an algebraic constraint,
once spherical and time harmonics have been introduced. This
allows for a quick determination of some electromagnetic field
configurations, such as the solutions describing a static magnetic
dipole or a static uniform magnetic field at infinity. We also
discuss the key wave equations for two classes of dust Universe
models. Specifically, we have a look at the inhomogeneous
Lema\^{i}tre-Tolman-Bondi (LTB) spacetimes
\cite{Kramer-Stephani-MacCallum-Herlt,Lemaitre,Tolman,Bondi}, and
the spatially homogeneous Kantowski-Sachs (KS) spacetime
\cite{K,KS,Kramer-Stephani-MacCallum-Herlt} in the spherical
symmetric case.

\section{Preliminaries}

Let us briefly review the basics of the 1+1+2 covariant approach
and introduce the required notation, referring the reader to
\cite{Clarkson-Barrett} for a more elaborate exposition. We adopt
units such that $8\pi G=1=c$ and Einstein's equations are
$G_{ab}=T_{ab}-\Lambda g_{ab}$.

While the 1+3 approach \cite{Ellis-vanElst} provides a threading
of spacetime into `time' and `space' with the means of a timelike
unit vector field $u^a$ ($u^au_a=-1$) , representing the
observers' 4-velocity, the 1+1+2 formalism in addition performs a
slicing of the `space' with the help of a spacelike unit vector
field $\n^a$, which is orthogonal to $u^a$:
$\n^a\n_a=1,~u^a\n_a=0$. The 1+3 projection tensor
$h_{a}^{~b}\equiv g_{a}^{~b}+u_au^b$ combined with $n^a$ gives
rise to a new projection tensor $N_{ab}$, \be \N_a^{~b}\equiv
h_a^{~b}-\n_a\n^b=g_{a}^{~b}+u_au^b-\n_a\n^b\ , \ee which projects
vectors orthogonal to $\n^a$ and $u^a$
($\n^a\N_{ab}=0=u^a\N_{ab}$) onto 2-surfaces ($\N_a^{~a}=2$) which
are referred to as the `sheets'. Its volume element, the
alternating Levi-Civita 2-tensor, is derived from the volume
element $\lc_{abc}\equiv u^d\eta_{dabc}$ for the observers'
restspaces by \be \lc_{ab}\equiv\lc_{abc}\n^c = u^d\eta_{dabc}n^c\
;\qquad \lc_{ab}\n^b=0=\lc_{(ab)}\ . \ee

The covariant 1+3 threading irreducibly splits any 4-vector  into
a scalar part parallel to $u^a$ and a 3-vector part orthogonal to
$u^a$. Furthermore, any second rank tensor is covariantly and
irreducibly split into scalar, 3-vector and projected symmetric
trace-free (PSTF) 3-tensor parts. The 1+1+2 slicing takes this
split further by irreducibly decomposing the 3-vectors and PSTF
3-tensors with respect to $n^a$. For example, any 3-vector
$\psi^a$ can now be irreducibly split into a scalar, $\Psi$, which
is the part of the vector parallel to $\n^a$, and a vector,
$\Psi^a$, lying in the sheet orthogonal to $\n^a$: \be \fl
\psi^a=\Psi\n^a+\Psi^{a},~~~\mbox{where}~~~\Psi\equiv \psi_a\n^a\
,~~~\mbox{and}~~~\Psi^{a}\equiv \N^{ab}\psi_b\equiv \psi^{\bar a}\
, \label{vector-decomp} \ee where we use a bar over an index to
denote projection with $\N_{ab}$. Similarly, any PSTF tensor,
$\psi_{ab}$, can now be split into scalar, vector and tensor
(which are PSTF with respect to $\n^a$) parts: \be
\psi_{ab}=\psi_{\langle
ab\rangle}=\Psi\bra{\n_a\n_b-\sfr12\N_{ab}}+2\Psi_{(a}\n_{b)}+\Psi_{{ab}}\
, \label{tensor-decomp} \ee where \ba
\Psi&\equiv &\n^a\n^b\psi_{ab}=-\N^{ab}\psi_{ab}\ ,\nonumber\\
\Psi_a&\equiv &\N_a^{~b}\n^c\psi_{bc}=\Psi_{\bar a}\ ,\nonumber\\
\Psi_{ab}&\equiv &
\bra{\N_{(a}^{~~c}\N_{b)}^{~~d}-\sfr{1}{2}\N_{ab}\N^{cd}}\psi_{cd}
\equiv \Psi_{\lb ab\rb}\ . \label{PSTF-TT}
\ea We use curly brackets to denote the PSTF with respect to
$\n^a$ part of a tensor. {Note that for 2nd-rank tensors in the
1+1+2 formalism `PSTF' is precisely equivalent to
`transverse-traceless'}.

In the 1+1+2 formalism three derivatives are used, which $u^a$ and
$\n^a$ define, for any object $\psi_{\cdots}^{~~\cdots}$: \ba \fl
\dot \psi_{a\cdots b}^{~~~~~c\cdots d}&\equiv & u^e
\nabla_e\psi_{a\cdots b}^{~~~~~c\cdots d}\ ,\label{dotdef}\\
\fl \hat \psi_{a\cdots b}^{~~~~~c\cdots d}&\equiv & \n^e
\sdel_e\psi_{a\cdots b}^{~~~~~c\cdots d} \equiv \n^e
h_e^{~j}h_a^{~f}\cdots h_b^{~g}h_h^{~c}\cdots h_i^{~d}\nabla_j
\psi_{f\cdots
  g}^{~~~~~h\cdots i}\ ,\label{hatdef}\\ \fl
\delta_e \psi_{a\cdots b}^{~~~~~c\cdots d}&\equiv &
\N_e^{~j}\N_a^{~f}\cdots \N_b^{~g}\N_h^{~c}\cdots\N_i^{~d}\sdel_j
\psi_{f\cdots
  g}^{~~~~~h\cdots i}\ .\label{deltadef}
\ea
The hat-derivative, analogously defined to the usual time (dot)
derivative, is the derivative along the vector field $\n^a$ in the
surfaces orthogonal to $u^a$. The $\delta$-derivative, defined by
equation~(\ref{deltadef}) similarly to the spatial derivative
$\sdel$, is a projected derivative on the sheet, with projection
on every free index.

With these definitions the spatial derivative of $\n^a$ orthogonal
to $u^a$ is decomposed as follows:
\be
\sdel_a\n_b=\n_a\hatn_b+\sfr12\phi \N_{ab}+\xi\lc_{ab}+\zeta_{ab}\
,
\ee
where
\ba
\hatn_a &\equiv &\n^c\sdel_c\n_a=\hat \n_a\ ,\\
\phi &\equiv &\delta_a \n^a\ ,\\
\xi &\equiv &\sfr12\lc^{ab}\delta_a\n_b\ ,\\
\zeta_{ab} &\equiv &\delta_{\lb a}\n_{b\rb}\ .
\ea
These  have a meaning analogous to the kinematical quantities
4-acceleration $\dot{u}^a$, expansion $\theta$, vorticity
$\omega^a$ and shear $\sigma_{ab}$ of the congruence $u^a$ in the
1+3 formalism, which are related via
\be
\nabla_au_b=-u_a\dot u_b+\sfr13\theta
h_{ab}+\lc_{abc}\omega^c+\sigma_{ab}\ ;
\ee
more precisely, travelling along $\n^a$, $\phi$ represents the
sheet expansion, $\zeta_{ab}$ is the shear of $\n^a$ (distortion
of the sheet), and $\hatn^a$ its acceleration, while $\xi$
represents a `twisting' of the sheet (the rotation of
$\n^a$~\cite{TM1}). The other derivative of $\n^a$ is its change
along $u^a$,
\be
\dot n_{a}=\udot u_a+\dotn_a~~~\mbox{where}~~~\dotn_a\equiv\dot
n_{\bar a} ~~~\mbox{and}~~~\udot=\n^a\uudot_a \ .
\ee
The new variables $\hatn_a$, $\phi$, $\xi$, $\zeta_{ab}$, $\udot$
and $\dotn_a$ are fundamental objects in the spacetime within the
1+1+2 approach since the spacetime geometry can be inferred from
them. They are treated on the same footing as the kinematical
variables of $u^a$ in the 1+3 approach.

The kinematical 1+3 quantities, the Weyl curvature and the
energy-momentum tensor need also to be split. For the
4-acceleration, vorticity and shear, one arrives at
\ba
\uudot^a&=&\udot \n^a+\udot^a\ ,\\
\omega^a&=&\Omega \n^a+\Omega^a\ ,\\
\sigma_{ab}&=&\Sigma\bra{\n_a\n_b-\sfr12\N_{ab}}+2\Sigma_{(a}\n_{b)}+\Sigma_{ab}\
,
\ea
while for the electric and magnetic Weyl tensors one gets
\ba
E_{ab}&=&{\cal E}\bra{\n_a\n_b-\sfr12\N_{ab}}+2{\cal E}_{(a}\n_{b)}+{\cal E}_{ab}\ ,\\
H_{ab}&=&{\cal H}\bra{\n_a\n_b-\sfr12\N_{ab}}+2{\cal
H}_{(a}\n_{b)}+{\cal H}_{ab}\ .
\ea
The fluid variables heat flux $q^a$ and anisotropic pressure
$\pi_{ab}$, inferred from the energy-momentum tensor $T_{ab}$
together with the energy density $\mu$ and the isotropic pressure
$p$, are split into
\ba
q^a&=&Q \n^a+Q^a\ ,\\
\pi_{ab}&=&\Pi\bra{\n_a\n_b-\sfr12\N_{ab}}+2\Pi_{(a}\n_{b)}+\Pi_{ab}\
.
\ea

Finally, the covariant 1+1+2 equations for all the above
introduced quantities may be obtained from the Bianchi and Ricci
identities for both the timelike unit vector field $u^a$
\emph{and} the spacelike unit vector field $n^a$.

\section{LRS class II spacetimes}
In this section, we discuss LRS class II spacetimes in terms of
the 1+1+2 formalism. The discussion follows in parts van Elst \&
Ellis \cite{vanElst-Ellis} but generalises their treatment of LRS
class II perfect fluids towards imperfect fluids employing a new,
somewhat streamlined notation.

The 1+1+2 formalism is ideally suited to study spacetimes which
exhibit local rotational symmetry (LRS) because they have a unique
preferred spatial direction at each point, defined covariantly,
for example, by an eigendirection of a degenerate rate of the
shear tensor field or by a vorticity vector field. This direction
constitutes a local axis of symmetry - all observations are
identical under rotations about it and are the same in all spatial
directions perpendicular to it~\cite{Ellis67,Stewart-Ellis}.
Hence, after a 1+1+2 split, if the spacelike unit vector field
$n^a$ $(n^an_a=1$, $n^au_a=0)$ is chosen parallel to the axis of
symmetry, all covariantly defined 1+1+2 vectors and tensors must
vanish due to the LRS symmetry. It follows that LRS spacetimes may
be covariantly characterised by scalar quantities (modulo
equations of state for the matter variables), namely
\be
\textbf{LRS}: \qquad \{\udot, \theta, \phi, \xi, \Sigma, \Omega,
{\cal E}, {\cal H},\mu, p, Q, \Pi, \Lambda \}\ . \label{LRSvars}
\ee
The variables \reff{LRSvars} thus \emph{fully} describe LRS
spacetimes and are what is solved for in the 1+1+2 approach. For
LRS class II, one requires in addition that the vorticity terms
vanish, $\Omega=0=\xi$, which further constrains the magnetic Weyl
curvature ${\cal H}$ to vanish. Thus, a substantially smaller set
of scalars describes LRS class II spacetimes:
\be
\textbf{LRS class II}: \qquad \{\udot, \theta, \phi, \Sigma, {\cal
E}, \mu, p, Q, \Pi, \Lambda \}\ .
\ee
These LRS class II quantities satisfy a bunch of covariant
evolution and/or propagation equations, obtained from the Bianchi
and Ricci identities for the unit vector fields $u^a$ and $n^a$,
respectively:

\textit{Propagation}:
\ba
\fl
\hat\phi&=&-\sfr12\phi^2+\bra{\sfr13\theta+\Sigma}\bra{\sfr23\theta-\Sigma}
    -\sfr23\bra{\mu+\Lambda}-\E -\sfr12\Pi\ ,\label{hatphinl}
\\\fl
\hat\Sigma-\sfr23\hat\theta&=&-\sfr32\phi\Sigma-Q\
,\label{Sigthetahat}
 \\\fl
\hat\E-\sfr13\hat\mu+\sfr12\hat\Pi&=&
    -\sfr32\phi\bra{\E+\sfr12\Pi}
    +\bra{\sfr12\Sigma-\sfr13\theta}Q\ ;
\ea

\textit{Evolution}:
\ba
\fl \dot\phi &=& -\bra{\Sigma-\sfr23\theta}\bra{\udot-\sfr12\phi}
+Q\ , \label{phidot}
\\ \fl
\dot\Sigma-\sfr23\dot\theta &=&
-\udot\phi+2\bra{\sfr13\theta-\sfr12\Sigma}^2
        +\sfr13\bra{\mu+3p-2\Lambda}-\E+\sfr12\Pi\ ,\label{Sigthetadot}
\\\fl
\dot\E -\sfr13\dot \mu+\sfr12\dot\Pi &=&
    +\bra{\sfr32\Sigma-\theta}\E
    +\sfr14\bra{\Sigma-\sfr23\theta}\Pi
    +\sfr12\phi Q
       -\sfr12\bra{\mu+p}\bra{\Sigma-\sfr23\theta}\ ; \label{edot}
\ea

\textit{Propagation/evolution}:
\ba
\fl \hat\udot-\dot\theta&=&-\bra{\udot+\phi}\udot+\sfr13\theta^2
    +\sfr32\Sigma^2 +\sfr12\bra{\mu+3p-2\Lambda}\ ,\label{Raychaudhuri}
\\
 \fl \dot\mu+\hat Q&=&-\theta\bra{\mu+p}-\bra{\phi+2\udot}Q -
    \sfr32\Sigma\Pi\ ,
\\ \fl \label{Qhat}
\dot Q+\hat
p+\hat\Pi&=&-\bra{\sfr32\phi+\udot}\Pi-\bra{\sfr43\theta+\Sigma} Q
    -\bra{\mu+p}\udot\ .
\ea
This system of equations extends the corresponding system in
\cite{vanElst-Ellis} (see section 6) to imperfect fluids.

Since the vorticity $\Omega$ vanishes, the unit vector field $u^a$
is hypersurface-orthogonal to the spacelike 3-surfaces whose
intrinsic curvature can be calculated from the Gauss equation for
$u^a$. For the intrinsic Ricci-curvature, one finds from this
\ba
^3R_{ab} &=& \bras{\sfr23\bra{\mu+\Lambda} +\E+\sfr12\Pi
  +\Sigma^2-\sfr13\theta \Sigma -\sfr29\theta^2}n_an_b \nonumber\\
  &&+ \bras{\sfr23\bra{\mu+\Lambda}-\sfr12\E-\sfr14\Pi
  +\sfr14\Sigma^2+\sfr16\theta \Sigma
  -\sfr29\theta^2}N_{ab}\ . \label{3ricci}
\ea
The intrinsic Ricci-scalar of the 3-surfaces is therefore
\be
^3R = 2\bras{\mu+\Lambda-\sfr13\theta^2+\sfr34\Sigma^2}\ .
\label{3scalar}
\ee
This relation constitutes the generalised Friedman equation.

On the other hand, the additional vanishing of the sheet
distortion, $\xi$, implies that the sheet is in this case a
genuine 2-surface. The Gauss equation for $n^a$ together with the
3-Ricci identities determine the 3-Ricci curvature tensor of the
spacelike 3-surfaces orthogonal to $u^a$ to be
\be
^3R_{ab} = -\bras{\hat \phi +\sfr12\phi^2}n_an_b -\bras{\sfr12
\hat
  \phi +\sfr12\phi^2 -K}N_{ab}\ , \label{3ric}
\ee
thus giving for the 3-Ricci scalar
\be
^3R =-2\bras{\hat \phi +\sfr34\phi^2 -K}\ . \label{3sca}
\ee
Here, $K$ is the Gaussian curvature of the sheet,
$^2R_{ab}=KN_{ab}$. Combining the last equation with
equations~\reff{3scalar} and \reff{hatphinl}, we can express the
Gaussian curvature $K$ in the form
\be
K = \sfr13\bra{\mu+\Lambda}-\E-\sfr12\Pi +\sfr14\phi^2
-\bra{\sfr13\theta-\sfr12\Sigma}^2\ . \label{gauss}
\ee
Evolution and propagation equations for $K$ can be calculated from
equations~\reff{hatphinl}-\reff{edot} to give
\ba
\dot K &=& -\bra{\sfr23\theta-\Sigma}K\ , \label{Kdot}\\
\hat K &=& -\phi K\ . \label{Khat}
\ea
Every scalar $\psi$ in LRS II has to satisfy the commutation
relation \be \hat{\dot \psi} - \dot{\hat \psi} = -\udot\dot \psi +
\bra{\sfr13\theta+\Sigma}\hat\psi\ . \ee We applied this
commutator relation to the Gaussian curvature $K$ in order to
check for consistency of the equations
\reff{hatphinl}--\reff{Qhat}, which is indeed guaranteed due to
equations \reff{Kdot} and \reff{Khat}.

From equation~\reff{Kdot} follows the neat result that whenever
the Gaussian curvature $K$ of the sheet is constant in time but
non-vanishing, that is, the sheets have either spherical or
hyperbolic geometry, the shear $\Sigma$ is proportional to the
expansion $\theta$: \be K \neq 0\quad \textrm{and}\quad \dot K =
0\quad \Longrightarrow \quad \Sigma = \sfr23\theta\ . \label{prop}
\ee As a matter of frame choice, it is always possible to pick an
observer with $\dot K =0$, corresponding to a `static' observer.
Adopting such a choice leads to great simplifications in the
equations, the caveat being that such a choice is often not the
most natural one, e.g., in Friedman-Lema\^{i}tre-Robertson-Walker
(FLRW) spacetimes. For example, taking a static observer to
describe spherically symmetric spacetimes, the constraint
\reff{gauss} allows for a covariant definition of the radial
parameter $r$ via \be r^{-2} =
\sfr13\bra{\mu+\Lambda}-\E-\sfr12\Pi +\sfr14\phi^2\ ; \qquad \dot
r = 0 = \delta_a r\ . \ee (A non-static observer has $\dot r \neq
0$). In situations with spherical symmetry it is convenient to
express the hat-derivative in terms of the radial parameter $r$.
If we associate an affine parameter $\rho$ with the
hat-derivative, we find using equation~\reff{Khat} \be \hat X =
\fr{dX}{d\rho} = \fr12 r\phi \fr{\partial X}{\partial r} + \hat
\tau \dot X\ , \ee where $\tau$ denotes proper time and $X$ may be
any scalar.

In summary, we have found that in the 1+1+2 formalism, all LRS II
spacetimes are covariantly described by the scalar equations
~\reff{hatphinl}--\reff{Qhat}. In this case, the sheets are
genuine 2-surfaces whose extrinsic geometry is determined by the
Gaussian curvature $K$, which can be computed from \reff{gauss}.
Furthermore, due to \reff{prop}, we stress that static spacetimes
whose sheet is either spherical or hyperbolic must have their
shear $\Sigma$ proportional to the expansion $\theta$; more
precisely, such \emph{imperfect} fluid spacetimes fulfill the
relation $\Sigma-\sfr23\theta=0$, generalising the well-known fact
that perfect fluid spacetimes within this subclass possess
vanishing shear $\Sigma$ and expansion $\theta$.

\section{Perturbations on LRS class II backgrounds}

In this section, we investigate scalar and electromagnetic
perturbations on LRS class II background spacetimes. The scalar
end electromagnetic fields are treated as test fields, i.e., they
do not affect the geometry of the background. Hence, the
Stewart-Walker Lemma~\cite{Stewart-Walker} ensures that these
fields are gauge-invariant. Our goal is the derivation of the
concomitant covariant and gauge-invariant wave equations which
describe these perturbations. These wave equations are the
generalisation of the well-known Regge-Wheeler equations from the
Schwarzschild case towards LRS class II spacetimes, yet written in
covariant notation.

\subsection{Commutation relations}

In order to achieve this aim, we will make frequent use of
commutation relations. All first order scalars $\psi$ have to
satisfy the following commutation relations between the different
derivatives introduced above:
\ba
\hat{\dot \psi}-\dot{\hat
  \psi}&=&-\udot\dot\psi+\bra{\sfr13\theta+\Sigma}\hat\psi\ ,\label{comm-dothat}
\\
\delta_a\dot\psi-\N_a^{~b}\bra{\delta_b\psi}^\cdot&=&-\sfr12
\bra{\Sigma-\sfr23\theta}\delta_a\psi\ , \label{comm-dotdel}
\\
\delta_a\hat\psi-\N_a^{~b}\bra{\delta_b\psi}^{\hat{}}
&=&+\sfr12\phi\delta_a\psi\ , \label{comm-hatdel}
\\
\delta_{[a}\delta_{b]}\psi&=&\, 0\ .\label{comm-deldel}
\ea
Analogously, the commutation relations for first order 2-vectors
$\psi_a$ read
\ba
\hat{\dot \psi}_{\bar a}-\dot{\hat \psi}_{\bar
  a}&=&-\udot\dot\psi_{\bar a}+\bra{\sfr13\theta+\Sigma}\hat\psi_{\bar a}
 \ ,\label{commv-un}
\\
\delta_a\dot\psi_b-\N_a^{~c}\N_b^{~d}\bra{\delta_c\psi_d}^\cdot&=&
-\sfr12 \bra{\Sigma-\sfr23\theta}\delta_a\psi_b\ ,
\label{commv-deldot}
\\
\delta_a\hat\psi_b-\N_a^{~c}\N_b^{~d}\bra{\delta_c\psi_d}^{\hat{}}&=&
+\sfr12\phi\delta_a\psi_b\ ,\label{commv-delhat}
\\
\delta_{[a}\delta_{b]}\psi_c &=& -K\psi_{[a}N_{b]c}\ .
  \label{commv1}
\ea
\subsection{Harmonics}

It is useful to expand all first-order perturbations in harmonics.
Note that all functions and relations below are defined in the
background only; we solely expand first-order variables, so
zeroth-order equations are sufficient.

In analogy to the spatial harmonics defined in
\cite{Dunsby-Bruni-Ellis}, we introduce dimensionless sheet
harmonic functions $Q$, defined on the background, as
eigenfunctions of the 2-dimensional Laplace-Beltrami operator such
that for positive, negative or vanishing curvature $K$
\be
\delta^2 Q = -\fr{k^2}{r^{2}} Q\ , \qquad\hat Q=0=\dot Q \qquad (0
\leq k^2\, \in\, \mathbb{R})\ .\label{SH}
\ee
The function $r$ is, up to an irrelevant constant, covariantly
defined by
\be
\fr{\hat r}{r}\, \equiv \fr12\phi\ , \qquad  \fr{\dot r}{r} \equiv
\fr13\theta-\fr12\Sigma\ , \qquad \delta_a r \equiv 0\ ,
\label{rdef}
\ee
which, in the light of equations \reff{Kdot}--\reff{Khat}, is
nothing but the covariant version of the common setting $K\equiv
\kappa /r^2$ with $\kappa = \pm 1$ for spherical and hyperbolic
2-geometry, respectively, where the constant is fixed by the
relation \reff{gauss}. However, the covariant definition
\reff{rdef} extends to the plane-symmetric case as well, where
there is no natural length scale. We can now expand any first
order scalar ${\psi}$ in terms of these functions formally as
\be
{\psi}=\sum_{k}{\psi}_{\Si}^{(k)} Q^{(k)} = {\psi}_{\Si} Q\ ,
\ee
where the sum over $k$  is implicit in the last equality. The
$\Si$ subscript reminds us that ${\psi}$ is a scalar, and that a
harmonic expansion has been made.

We also need to expand vectors in  harmonics. We therefore define
the \emph{even} (electric) parity vector  harmonics as \be \fl
Q_a^{(k)}=r\delta_a Q^{(k)} ~~~\Rightarrow ~~~\hat Q_{\bar
a}=0=\dot Q_{\bar a}\ ,~~~\delta^2Q_a=\bra{1-k^2}r^{-2}Q_a\ ; \ee
where the $(k)$ superscript is implicit, and we define \emph{odd}
(magnetic) parity vector harmonics as \be \fl \bar
Q_a^{(k)}=r\lc_{ab}\delta^b Q^{(k)}~~~\Rightarrow
~~~\hat{\bar{Q}}_{\bar a}=0=\dot{\bar{Q}}_{\bar a}\
,~~~\delta^2\bar Q_a=\bra{1-k^2}r^{-2}\bar Q_a\ . \ee Note that
$\bar Q_a=\lc_{ab}Q^b\Leftrightarrow Q_a=-\lc_{ab}\bar Q^b$, so
that $\lc_{ab}$ is a parity operator. The crucial difference
between these two types of vector harmonics is that $\bar Q_a$ is
solenoidal, so \be \delta^a\bar Q_a=0\ , \ee while \be
\delta^aQ_a=-k^2r^{-1} Q\ . \ee Note also that \be
\lc_{ab}\delta^a Q^b=0\ ,~~~\mbox{and}~~~\lc_{ab}\delta^a\bar
Q^b=+k^2r^{-1} Q\ . \ee The harmonics are orthogonal: $Q^a\bar
Q_a=0$ (for each $k$), which implies that any first-order vector
${\psi}_a$ can now be written \be {\psi}_a=\sum_{k}
{\psi}^{(k)}_{\Vi} Q_a^{(k)}+\bar {\psi}^{(k)}_{\Vi}\bar
Q_a^{(k)}={\psi}_{\Vi} Q_a+\bar {\psi}_{\Vi}\bar Q_a\ . \ee Again,
we implicitly assume a sum over $k$ in the last equality, and the
$\Vi$ subscript reminds us that ${\psi}_a$ is a vector expanded in
harmonics.

We like to point out that the harmonics introduced here naturally
generalise the spherical harmonics used in
\cite{Clarkson-Barrett}. In particular, the various formulae for
scalar and vector spherical harmonics stated in
\cite{Clarkson-Barrett} also hold for our generalised harmonics.

\subsection{Scalar perturbations}

Let us consider perturbations of LRS spacetimes due to a
\textit{massive} scalar field, $\psi$, with mass $M$, whose
equation of motion is the Klein-Gordon equation,
\be
\bra{g^{ab}\nabla_a\nabla_b +M^2} \psi
=\bra{\nabla^a\nabla_a+M^2}\psi = 0\ . \label{kgm}
\ee
We investigate this simple equation first in an arbitrary
spacetime and specialising to LRS II spacetime afterwards. The
corresponding 1+3 equation is easily derived using
$g^{ab}=h^{ab}-u^a u^b$ and the following expression for the
spatial Laplacian of a scalar field,
\be
\sdel^a\sdel_a \psi = h^{ab}\sdel_a\sdel_b \psi =
h^{ab}\nabla_a\nabla_b \psi + \theta\dot\psi\ , \label{slapm}
\ee
and leads to the wave equation
\be
\ddot\psi -\sdel^a\sdel_a \psi +\theta\dot\psi- \dot
u^a\sdel_a\psi +M^2\psi = 0\ . \label{31wm}
\ee
Thus, for a general spacetime, the evolution of a scalar field is
affected by effects of expansion and acceleration, as measured by
the fundamental observer.

The 1+1+2 form of equation~\reff{31wm} is readily obtained using
$h^{ab}=N^{ab}+n^a n^b$ as well as the relation \be
\delta^a\delta_a \psi = N^{ab}\delta_a\delta_b \psi =
N^{ab}\sdel_a\sdel_b \psi - \phi\hat\psi\ . \label{Slapm} \ee The
resulting equation reads as \be \ddot\psi -\hat{\hat \psi}
+\theta\dot\psi- (\udot+\phi)\hat\psi + (a^a - \udot^a
-\delta^a)\delta_a \psi +M^2\psi = 0\ . \label{112wm} \ee
Equation~\reff{112wm} is the fully split Klein-Gordon equation for
a massive scalar field in an arbitrary spacetime, given in a
covariant and gauge-invariant fashion. For LRS spacetimes, the
wave equation \reff{112wm} closes, yielding
\be \ddot\psi -\hat{\hat
\psi} +\theta\dot\psi- (\udot+\phi)\hat\psi + (M^2 -\delta^2) \psi
= 0\ . \label{112lrs} \ee
 We emphasise that this is true for \textit{all} LRS spacetimes.
Equation~\reff{112lrs} constitutes the generalised Regge-Wheeler
 equation for scalar perturbations on all LRS  background
 spacetimes. While the generalised Regge-Wheeler equation
 \reff{112lrs} is not affected by rotational effects at all, it is
 in general sensitive to the observer's acceleration $\udot$, the spacetime
 expansion $\theta$, the sheet expansion $\phi$, as well as the
 mass $M$ of the perturbation field $\psi$.

 For LRS class II spacetimes, it is sometimes
 convenient to rescale the scalar field $\psi$ according to
 $\psi\equiv r^{-1}\Psi$, where the function $r$ is defined via
 \reff{rdef}. In terms of the rescaled field $\Psi$, the
 Regge-Wheeler equation \reff{112lrs} reads
\be \fl
\ddot\Psi -\hat{\hat \Psi} +\bra{\Sigma+\sfr13\theta}\dot\Psi-
\udot\hat\Psi + \bras{M^2 -\E -\sfr16\bra{\mu-3p+4\Lambda}
-\delta^2} \Psi = 0\ . \label{RW2}
\ee
This form of the Regge-Wheeler equation has several advantages
over \reff{112lrs}: firstly, it allows to introduce the Gaussian
curvature $K$ at the expense of the Weyl curvature $\E$, for
example, and secondly, it simplifies in vacuo. We emphasise that
this form is \emph{generic}: we will show in the next section that
electromagnetic perturbations are described covariantly by the
same equation but having a different potential [that is, the term
in square brackets in \reff{RW2}] once appropriate harmonics have
been used to get rid of the sheet Laplacian $\delta^2$.

\subsection{Electromagnetic perturbations}

In accordance with \reff{vector-decomp}, the electric, magnetic
and current 3-vector fields are irreducibly decomposed into scalar and
2-vector parts as
\be
    E^a = \Es n^a + \Es^a\ , \quad  B^a = \Bs n^a + \Bs^a\ ,\quad J^a = \Js n^a + \Js^a\
    .
\ee Thus, Maxwell's equations for electromagnetic test fields,
sourced by the total electric charge $\rho_{\mathrm e}$ and the
currents $\Js$ and $\Js_a$, respectively, on a LRS class II
geometry become in the sheet approach (the splitting for arbitrary
spacetimes is displayed in \cite{EMsignature}): \ba \fl \hat
\Es+\delta_a \Es^a &=& -\phi \Es+ \rho_{\mathrm e} \
,\label{m1lrs} \\ \fl \hat \Bs+\delta_a \Bs^a &=& -\phi \Bs\
,\label{m2lrs} \\ \fl \dot \Es-\lc_{ab}\delta^a \Bs^b &=&
+\bra{\Sigma-\sfr23\theta}\Es -\Js\ ,\label{m3lrs}\\ \fl \dot
\Bs+\lc_{ab}\delta^a \Es^b &=& +\bra{\Sigma-\sfr23\theta}\Es\
,\label{m4lrs}\\ \fl \dot \Es_{\bar a}+\lc_{ab}\bra{\hat
\Bs^b-\delta^b \Bs} &=& -\bra{\sfr12\phi+\udot}\lc_{ab}\Bs^b-
\bra{\sfr12\Sigma+\sfr23\theta}\Es_a -\Js_a\ ,\label{m5lrs}\\
\fl\dot \Bs_{\bar a}-\lc_{ab}\bra{\hat \Es^b-\delta^b \Es} &=&
+\bra{\sfr12\phi+\udot}\lc_{ab}\Es^b -
\bra{\sfr12\Sigma+\sfr23\theta}\Bs_a \ .\label{m6lrs}
 \ea
Note that $\lc_{ab}$ is a parity operator,
$\lc_{ac}\lc^{c}_{~b}=-N_{ab}$, thus we can switch parity between
2-vectors $X_a$ and $Y_a$ via \be X_a = \lc_{ab}Y^b
\Longleftrightarrow Y_a = -\lc_{ab}X^b\ . \label{parity} \ee Using
this relation, the parity-reversed form of equations~\reff{m5lrs}
and \reff{m6lrs} read \ba \fl \hat \Es_{\bar a}+\lc_{ab}\dot
\Bs^b-\delta_a\Es &=& -\bra{\sfr12\phi+\udot}\Es_a-
\bra{\sfr12\Sigma+\sfr23\theta}\lc_{ab}\Bs^b\ ,\label{parm5lrs}\\
\fl \hat \Bs_{\bar a}-\lc_{ab}\dot \Es^b-\delta_a\Bs &=&
+\bra{\sfr12\phi+\udot}\Bs_a -
\bra{\sfr12\Sigma+\sfr23\theta}\lc_{ab}\Es^b +\lc_{ab}\Js^b\
.\label{parm6lrs}
 \ea
From Maxwell's equations, one deduces (by a somewhat tedious
calculation exploiting the earlier displayed commutation relations
as well as the above parity-reversed equations) the following wave
equations for the electromagnetic fields along the distinguished
direction $n_a$: \ba
 \ddot \Es &-\hat{\hat \Es} - \bra{{\cal A} + 2\phi} \hat \Es -\bra{\Sigma-\sfr53\theta}\dot
 \Es \nonumber\\
  &- \bras{\delta^2  + \sfr12 \phi^2 -2\E +\bra{\sfr13\theta+\Sigma}\bra{\sfr32\Sigma-\theta} - \sfr13\bra{\mu-3p+4\Lambda}}\Es
  \nonumber \\
   &= \hat \rho_{\mathrm e}+ \dot\Js + \bra{\phi+\udot}\rho_{\mathrm e}+ \theta\Js\ ,
 \label{Ecosc1}\\
  \ddot \Bs &-\hat{\hat \Bs} - \bra{{\cal A} + 2\phi} \hat \Bs -\bra{\Sigma-\sfr53\theta}\dot \Bs  \nonumber \\
  &- \bras{\delta^2 + \sfr12 \phi^2 -2\E +\bra{\sfr13\theta+\Sigma}\bra{\sfr32\Sigma-\theta}
  - \sfr13\bra{\mu-3p+4\Lambda}}\Bs \nonumber \\
  &=0\ .\label{Bcosc2}
\ea
Equations~\reff{Ecosc1}--\reff{Bcosc2} are the generalisation of
the famous Regge-Wheeler equation, for electromagnetic
perturbations on the Schwarzschild background, towards LRS class
II spacetimes, although written in covariant manner. We emphasise
that  $\Es$ and $\Bs$ decouple from each other and that in the
absence of sources the equations are identical closed wave
equations.

It will turn out advantageous to rescale the fields and the
sources under consideration similarly as above for Klein-Gordon
fields; that is, we define
\be
    \Es \equiv r^{-2}E\ , \quad \Bs \equiv r^{-2}B\ , \quad \rho_{\mathrm e}\equiv r^{-2}\varrho_{\mathrm e}\ ,
   \quad \Js \equiv r^{-2}J\ , \label{EBrescale}
\ee
and substitute into the wave equations
\reff{Ecosc1}--\reff{Bcosc2} to obtain
\ba
 \ddot E &-&\hat{\hat E} -{\cal A} \hat{E} +\bra{\Sigma+\sfr13\theta}\dot E
 -\bras{\delta^2 + \bra{\Sigma-\sfr23\theta}\bra{2\Sigma+\sfr16\theta}} E \nonumber \\
 &=& \hat \varrho_{\mathrm e}+ \dot J +\udot \varrho_{\mathrm e}+ \bra{\Sigma + \sfr13\theta}J\ ,
 \label{Eregge}\\
 \ddot B &-&\hat{\hat B} - {\cal A} \hat{ B}
 +\bra{\Sigma+\sfr13\theta}\dot B
 -\bras{\delta^2+ \bra{\Sigma-\sfr23\theta}\bra{2\Sigma+\sfr16\theta}}B =0.
  \label{Bregge}
\ea
Again, if we neglect the source terms, the equations become
identical closed wave equations. Moreover, the equations
\reff{Eregge}--\reff{Bregge} are of the same form as equation
\reff{RW2} for the Klein-Gordon field, the only difference being a
slightly altered potential term.

The wave equations for the electromagnetic perturbations lying in
the sheet are derived analogously and read \ba \fl \ddot \Es_{\bar
a} - \hat{\hat \Es_{\bar a}} &-& \bra{3{\cal A} + \phi} \hat
\Es_{\bar a} -\bra{\Sigma-\sfr53\theta}\dot \Es_{\bar a}
 \nonumber \\ \fl
 &+& \bras{\sfr14\phi^2 - \E -\udot\bra{\udot+\phi}
  + \sfr29\theta^2 -\sfr23\theta\Sigma -\sfr74\Sigma^2 + \sfr13\bra{\mu-3p+4\Lambda}-\delta^2}\Es_{a}
 \nonumber \\ \fl
  &=&\bra{\phi - 2\udot}\delta_a\Es +3\Sigma\lc_{ab}\hat \Bs^b +\bra{3\udot\Sigma-Q +\hat
  \theta-\dot\udot}\lc_{ab}\Bs^b \nonumber \\ \fl
  &&-\delta_a\rho_{\mathrm e}-\dot\Js_{\bar a}-\bra{\theta-\sfr32\Sigma}\Js_a\ ,
 \label{Earegge}\\ \fl
 \ddot \Bs_{\bar a} -\hat{\hat \Bs_{\bar a}} &-& \bra{3{\cal A} + \phi} \hat \Bs_{\bar a} -\bra{\Sigma-\sfr53\theta}\dot \Bs_{\bar a}
 \nonumber\\ \fl
  &+& \bras{\sfr14 \phi^2 -\E -\udot\bra{\udot+\phi}
  + \sfr29\theta^2 -\sfr23\theta\Sigma -\sfr74\Sigma^2 +
  \sfr13\bra{\mu-3p+4\Lambda}-\delta^2}\Bs_{a}\nonumber
  \\ \fl
&=&\bra{\phi-2\udot}\delta_a\Bs -3\Sigma\lc_{ab}\hat \Es^b
-\bra{3\udot\Sigma-Q +\hat \theta-\dot\udot}\lc_{ab}\Es^b
\nonumber \\ \fl
 &&+\lc_{ab}\bras{\delta^b\!\Js
-\hat\Js^b-\bra{\sfr12\phi+2\udot}\Js^b} \ . \label{Baregge}
\ea
In contrast to equations~\reff{Ecosc1}--\reff{Bcosc2} , these do
not decouple--not even in the absence of sources. For example, in
addition to the source terms, the magnetic 2-vector field $\Bs_a$
gives rise to forcing terms for the electric 2-vector field
$\Es_a$, which is also forced by the `radial' electric field.

However, it turns out that \emph{in the absence of sources}
knowledge of the `radial' part $\Es$ (or $\Bs$) of the
perturbations, e.g., by solving the concomitant Regge-Wheeler
equation \reff{Eregge}, suffices to completely determine the
electromagnetic perturbations. To see this, we expand all
perturbations into harmonics and decompose the governing Maxwell's
equations into their harmonic components. (For details, we refer
the reader to Ref.~\cite{Clarkson-Barrett}, Section E). From the
scalar equations~\reff{m1lrs}--\reff{m4lrs} we find for each fixed
harmonic index $k$
\ba
\hat \Es_{\Si}   +\phi \Es_{\Si} &=& +\sfr{k^2}{r}\Es_{\Vi} \ , \label{ms1}\\
\hat \Bs_{\Si}   +\phi \Bs_{\Si} &=& +\sfr{k^2}{r}\Bs_{\Vi} \ , \label{ms2}\\
\dot \Es_{\Si} -\bra{\Sigma-\sfr23\theta}\Es_{\Si}  &=& +\sfr{k^2}{r}\bar \Bs_{\Vi} \ , \label{ms3}\\
\dot \Bs_{\Si} -\bra{\Sigma-\sfr23\theta}\Bs_{\Si}  &=&
-\sfr{k^2}{r}\bar \Es_{\Vi} \ . \label{ms4}
\ea
It is thus obvious that a solution for the `radial' fields $\Es$
and $\Bs$ determines the sheet fields $\Es_a$ and $\Bs_a$,
respectively.

The 2-vector equations~\reff{m5lrs} and \reff{m6lrs}, or
~\reff{parm5lrs} and ~\reff{parm6lrs}, respectively,  split into
two parts; the first part is the odd parity part, given by
\ba
\dot {\bar \Es}_{\Vi} + \hat \Bs_{\Vi} - \sfr1r \Bs_{\Si} &=&
-\bra{\sfr12 \phi + \udot} \Bs_{\Vi} -
\bra{\sfr12\Sigma+\sfr23\theta}
\bar \Es_{\Vi}
\ , \label{ms5}\\
\dot {\bar \Bs}_{\Vi} - \hat \Es_{\Vi} + \sfr1r \Es_{\Si} &=&
+\bra{\sfr12 \phi + \udot} \Es_{\Vi} -
\bra{\sfr12\Sigma+\sfr23\theta}
\bar \Bs_{\Vi} \ , \label{ms6}
\ea
and the second part is the even parity part, given by
\ba
\dot \Es_{\Vi} - \hat{\bar \Bs}_{\Vi} &=& +\bra{\sfr12 \phi +
\udot}\bar \Bs_{\Vi}
- \bra{\sfr12\Sigma+\sfr23\theta} \Es_{\Vi}\ , \label{ms7}\\
\dot \Bs_{\Vi} + \hat{\bar \Es}_{\Vi} &=& -\bra{\sfr12 \phi +
\udot}\bar \Es_{\Vi} - \bra{\sfr12\Sigma+\sfr23\theta}\Bs_{\Vi}\ .
\label{ms8}
\ea
We remark that the even parity equations are redundant since
propagating the constraints \reff{ms3}--\reff{ms4} and inserting
them into \reff{ms7} and \reff{ms8} just yields back
equation~\reff{ms1} and \reff{ms2}. On the other hand, the odd
parity equations are not implied by the scalar ones and have been
used in the derivation of the generalised Regge-Wheeler equation
\reff{Ecosc1}. If one eliminates $\bar \Es_{\Vi}$ and $\bar
\Bs_{\Vi}$ from equations \reff{ms7}--\reff{ms8} using the
constraints \reff{ms3}--\reff{ms4}, it becomes obvious that the
electromagnetic perturbations fall into two distinct classes whose
equations decouple from each other, namely:
\ba
    \textrm{\emph{polar} perturbations}: \qquad \{\Es_{\Si},\Es_{\Vi},\bar
    \Bs_{\Vi}\}\ ,\label{polar}\\
    \textrm{\emph{axial} perturbations}: \qquad \{\Bs_{\Si},\Bs_{\Vi},\bar
    \Es_{\Vi}\}\ .\label{axial}
\ea
Moreover, the resulting equations involving either $\Es_{\Si}$ and
$\Es_{\Vi}$ or $\Bs_{\Si}$ and $\Bs_{\Vi}$ are identical.

\section{Examples}
We have shown that scalar and vector perturbations in LRS class II
spacetimes are governed by simple master equations. We shall now
discuss these master equations in some specific spacetimes. First,
we shall recover the well known Regge-Wheeler equations for the
Schwarzschild spacetime, which we then generalise to the Vaidya
radiation spacetime. We then discuss solutions from the Lema\^\i
tre-Tolman-Bondi and Kantowski-Sachs families.

\subsection{The Schwarzschild spacetime}
\subsubsection{The background}

Schwarzschild spacetime is fully determined by any two of three
non-zero scalar functions ${\phi\ ,\
  \udot\ ,\ \cal E}$.
These functions obey the background equations
\ba
\hat\phi &=&-\sfr12\phi^2+{\udot\phi}\ ,\label{phihatrad}\\
\hat{\udot}&=&-\udot\bra{\phi+\udot}\ ;\label{acchatrad}
\ea
together with the constraint
\be
{\cal E}+\udot\phi=0\ .\label{udotbackground}
\ee
The parametric solutions for these variables
are~\cite{Clarkson-Barrett}
\ba
{\cal E}&=&-\fr{2m}{r^3}\ ,\\
\phi &=&\fr2r\sqrt{1-\fr{2m}{r}}\ ,\\
\udot&=&\fr{m}{r^2}\bra{1-\fr{2m}{r}}^{-1/2}\ .\label{backsols(x)}
\ea
These form a one-parameter family of solutions, parameterised by
the constant $m$, which is just the Schwarzschild mass. The
(exterior) Schwarzschild solution is given for $2m<r<\infty$.

\subsubsection{The perturbations}
In order to make the connection with the standard Regge-Wheeler
equations \cite{Nollert}, it is instructive to contrast the found
covariant and gauge-invariant wave equations for scalar and
electromagnetic perturbations with the ones governing
gravitational perturbations. The fundamental object for the study
of gravitational perturbations on a Schwarzschild background is
the Regge-Wheeler tensor $W_{ab}$~\cite{Clarkson-Barrett}, defined
by
\be
W_{ab}=\sfr12\phi r^2\zeta_{ab}-\sfr13r^2\E^{-1}\delta_{\lb
  a}\delta_{b\rb}\E\ ,\label{Wab}
\ee a dimensionless, gauge-invariant, transverse-traceless tensor
which obeys the  wave equation \be \ddot W_{\lb ab\rb}-\hat{\hat
W}_{\lb ab\rb}-\udot{\hat W}_{\lb ab\rb} +\phi^2 W_{ab}-\delta^2
W_{ab} =0\ .\label{RWtensorwave} \ee If one expands equation
~(\ref{RWtensorwave}) into spherical tensor harmonics, the odd and
even parity parts of equation~(\ref{RWtensorwave}) both become \be
\ddot W_{\Ti}-\hat{\hat W}_{\Ti}-\udot\hat
W_{\Ti}+\bras{\fr{L}{r^2}+3\E}W_{\Ti}=0\ ,\label{RW} \ee where
$W_{\Ti}=W_{\Ti}^{(\ell)}$ are the tensor harmonic components of
$W_{ab}$, $\ell=1,2,...$ and $L=\ell(\ell+1)=k^2$. It turns out
that equation~(\ref{RW}) is actually the Regge-Wheeler
equation~\cite{RW} when written in appropriate coordinates.
Converting from $\rho$, the affine parameter associated with the
hat-derivative, to the parameter $r$, $\rho\rightarrow r$, and
then to the `tortoise' coordinate of Regge and Wheeler, \be
r_*=r+2m\ln\bra{\fr{r}{2m}-1}\ ,\label{tortoise} \ee and also
introducing the Schwarzschild time via
$d\tau=\sqrt{1-\sfr{2m}{r}}\, dt$, we find that~(\ref{RW})
becomes: \be \fl \bra{-\fr{\partial^2}{\partial
t^2}+\fr{\partial^2}{\partial r_*^2}+\mathscr{V}_{T}}W_{\Ti}=0 \
,\qquad
\mathscr{V}_{T}=\bra{1-\fr{2m}{r}}\bras{\fr{L}{r^2}-\fr{6m}{r^3}}\
, \label{schroed} \ee where $\mathscr{V}_{T}$ is the Regge-Wheeler
potential for gravitational perturbations.

It is now straightforward to obtain the familiar Regge-Wheeler
equations for the case of scalar and electromagnetic
perturbations. From equation~\reff{RW2} one finds immediately for
a massive scalar field $\psi$ with mass $M$, $\psi=r^{-1}\Psi$,
that \be \ddot\Psi -\hat{\hat \Psi} -\udot\hat\Psi -\bras{\E-M^2 +
\delta^2} \Psi =0\ . \label{112r} \ee Introducing scalar spherical
harmonics and performing as above one readily gets \be \fl
\bra{-\fr{\partial^2}{\partial t^2}+\fr{\partial^2}{\partial
r_*^2}+\mathscr{V}_{S}}\Psi_{\Si}=0 \ ,\qquad \
\mathscr{V}_{S}=\bra{1-\fr{2m}{r}}\bras{\fr{L}{r^2}+\fr{2m}{r^3}+M^2}\
, \label{RWscalar} \ee where $\Psi_{\Si}$ is a scalar harmonic
component of $\Psi$ and $\mathscr{V}_S$ is the Regge-Wheeler
potential for scalar perturbations. This equation was originally
derived in \cite{PriceI} employing the Newman-Penrose formalism.
Notice the striking similarity between the Regge-Wheeler potential
for the massless scalar field perturbations, $\mathscr V_{S}$, and
the potential for gravitational perturbations,
$\mathscr{V}_{T}$~\cite{Nollert}.

The electromagnetic case is even simpler. For this case, the
covariant Regge-Wheeler equation~\reff{Eregge} for the electric
field perturbation $\Es$, $\Es=r^{-2}E$, reduces in the
Schwarzschild background to
\be
\ddot E -\hat{\hat E} -\udot\hat E - \delta^2 E =0\ . \label{rwem}
\ee
Adopting once more scalar spherical harmonics and proceeding as
before leads to
\be
\bra{-\fr{\partial^2}{\partial t^2}+\fr{\partial^2}{\partial
r_*^2}+\mathscr{V}_{V}}E_{\Si}=0 \ ,\quad
\mathscr{V}_{V}=\bra{1-\fr{2m}{r}}\bras{\fr{L}{r^2}}\ ,
\label{RWem}
\ee
where $\mathscr V_V$ is the Regge-Wheeler potential for
electromagnetic perturbations (the magnetic case is completely
analogous). The Regge-Wheeler equation for radial electric
perturbations~\reff{RWem}, and its magnetic counterpart, was
originally derived in \cite{PriceII} employing the Newman-Penrose
formalism.

It is remarkable that the (source-free) electromagnetic
perturbations can also be described by a linear first order system
of ODE's after expanding Maxwell's equations into spherical
\emph{and} time harmonics~\cite{Clarkson-Barrett} (possible
because the background is static). The time derivatives of first
order quantities are decomposed into their Fourier components by
assuming an $e^{i\omega\tau}$ time dependence for the first order
variables; factors of $i\omega$ just represent time derivatives,
$d/d\tau$. Note that
\be
\hat\omega=-\udot\omega~~~\Rightarrow~~~\omega=
\sigma\bra{1-\frac{2m}{r}}^{-1/2}=\frac{2\sigma}{\phi r}\
,\label{omdef}
\ee
arising from the commutation relation between the dot- and
hat-derivatives. The harmonic function $\omega$ is defined with
respect to proper time  $\tau$ of observers moving along $u^a$,
while $\sigma$ is the {\em constant} harmonic index associated
with time $t$ of observers at infinity. They are related by
$\omega\tau = \sigma t$. Inserting time harmonics into equations
\reff{ms1}--\reff{ms6} one finds that equations
\reff{ms3}--\reff{ms4} turn into constraints for the odd modes
$\bar \Es_{\Vi}$ and $\bar \Bs_{\Vi}$, respectively,
\be
    \bar \Bs_{\Vi}= \fr{i\omega r}{L}\Es_{\Si}\ , \qquad \bar \Es_{\Vi}= -\fr{i\omega
    r}{L}\Bs_{\Si}\ ,\label{constraints}
\ee
which may subsequently be eliminated from equations
\reff{ms5}--\reff{ms6}. The final result is a linear first order
system of ODE's,
\be \mathbf{\hat Y} = \mathbf{A} \mathbf{Y}\ ,\label{matrixeq} \ee
where the matrix $\mathbf{A}$ is given as
\be \mathbf{A} =
\left(
\begin{array}{cc}
    -\phi & \fr Lr \\
    \fr1r - \fr{r\omega^2}{L} & -\fr12\phi -\udot
\end{array}
\right) \label{matrix}
\ee
and the perturbation vector $\mathbf{Y}$ is defined in the case of
polar and axial perturbations by
\be
\mathbf{Y}_{\mathrm{polar}}= \left (
\begin{array}{c}
  \Es_{\Si} \\ \Es_{\Vi}
\end{array}
\right) \qquad \textrm{and} \qquad \mathbf{Y}_{\mathrm{axial}}=
\left (
\begin{array}{c}
  \Bs_{\Si} \\ \Bs_{\Vi}
\end{array}
\right)\ ,
\ee
respectively. We emphasise the observation that in a stationary
situation, e.g. $\omega=0$, the constraints \reff{constraints}
imply $\bar \Es_{\Vi} =
\bar \Bs_{\Vi} = 0$ in the absence of charges, which means that
electromagnetic perturbations cannot have a solenoidal part in the
sheet. A further implication is that a pure magnetic field can
only exist on the Schwarzschild geometry if it has no solenoidal
contribution in the sheet and is stationary and/or sheet-like. In
the stationary case, the system \reff{matrixeq} can be solved
analytically. For example, picking $L=2$ in \reff{matrix} easily
yields the well-known solutions of a static magnetic dipole or a
static uniform magnetic field at infinity (compare also
\cite{Wald,Hanni-Ruffini,Sonego-Abramowicz}).

\subsection{Vaidya Spacetime}

\subsubsection{The background}

The energy-momentum tensor of Vaidya's radiating sphere spacetime
is that of a radiation fluid, e.g., \be T_{ab} = \mu k_a k_b\ ,
\qquad k_ak^a = 0\ . \ee A convenient form of its metric for the
case of outgoing radiation is given in terms of retarded time
$u=t-r_*$ by
\be
    g_{ab}=-\bra{1-\fr{2m(u)}{r}}du^2 -2dudr+r^2d\Omega^2\ ,
    \label{metric}
\ee
where $m(u)$ is a arbitrary non-increasing function, $m'(u)\equiv
dm(u)/du\leq 0$; if $m(u)$ is constant, the metric is equivalent
to the Schwarzschild metric. Thus, the Vaidya spacetime describes
a collapsing star that irradiates a part of its mass, where the
total power output measured at infinity is given by $-m'(u)=4\pi
r^2\mu$
\cite{Vaidya2,Vaidya3,Raychaudhuri,Israel,LSM,Poisson-Israel}.

A covariant description of the Vaidya spacetime is obtained as
follows. Spherical symmetry requires the null-vector $k_a$ to be
of the form $k_a = u_a \pm n_a$, where a plus (minus) corresponds
to outflowing (inflowing) radiation. Thus, a static observer will
encounter the radiation fluid as an imperfect one, with the
energy-density $\mu$, the isotropic pressure $p=\sfr13 \mu$, the
radial heat-flux $Q=\pm\mu$ and the trace-free part of the
anisotropic sheet pressure $\Pi=\sfr23 \mu$. Working in the static
frame $(\dot r =0)$, the non-stationary spherically symmetric
Vaidya spacetime in the case of outgoing radiation is described
covariantly by the following set of equations:
\ba
\hat{\udot}-\dot\theta
&=& -\udot\bra{\udot+ \phi} + \theta\bra{\theta-\phi}\ , \\
\label{Ah} \hat{\phi} &=& -\phi\bra{\sfr12 \phi- \udot-2\theta}\ ,
\\\label{Ph} \dot{\phi} &=& - \theta\phi\ , \\\label{Pd}
\hat{\theta} + \dot{\theta} &=&-\theta\bra{\sfr12 \phi
+3\theta+3\udot}\ . \label{mhd}
\ea
The corresponding constraints, implied by our frame choice , are
\ba
K &\equiv & r^{-2} = \sfr14\phi^2-\E\ ,\\
\Sigma &=& \sfr23\theta\ , \\
\mu &=& -\theta\phi = 3p = Q = \sfr32\Pi\ , \\
\E &=& \mu-\udot\phi\ .
\ea
Note that the equations governing the Vaidya spacetime reduce to
the ones describing  Schwarzschild spacetime [see
\reff{phihatrad}--\reff{udotbackground}] when the expansion
$\theta$ vanishes.

In order to solve the equations, we find it useful to express our
derivatives in terms of coordinates $\{r,u\}$, where $r$ is the
radial parameter and $u$ labels time. In particular, we have
$\rho=\rho(r,u)$ and $\tau=\tau(r,u)$ and the derivatives become
for an arbitrary scalar $X$:
\ba
    \hat X =\fr{r\phi}{2}\fr{\partial X}{\partial r} + \hat u\fr{\partial X}{\partial
    u}\ ,\\
    \dot X = \dot u\fr{\partial X}{\partial u}\ .
\ea
One neat possibility to define the coordinate $u$ is through the
relations
\be
    \dot u = -\hat u = \fr{1}{\hat r} = \fr{2}{r\phi}\ ,
\ee
which is compatible with the commutator relation
\reff{comm-dothat} and identifies $u$ with the $u$-coordinate used
in the metric \reff{metric}, as will become clear in an instant.
Upon switching to the coordinates $\{r,u\}$ and putting  $\udot
\equiv \udot_{\mathsf{Sch}}-\theta$ [as suggested by equations
\reff{Ah} and \reff{mhd}], the system of equations transforms into
\ba
\bra{\sfr14r^2\phi^2\partial_r-\partial_u}\udot_{\mathsf{Sch}}&=&
-\sfr12r\phi\bras{\udot_{\mathsf{Sch}}\bra{\udot_{\mathsf{Sch}}+
\phi+\theta} +\sfr12\phi\theta}\ , \label{pASh}\\
 r\,\partial_r\phi &=& -\phi+2\udot_{\mathsf{Sch}}\ ,\label{pPSh}
\\ 2\,\partial_u \phi &=& -r \theta\phi^2 \ , \label{pPSd}\\
r\phi\,\partial_r\theta &=&-\theta\bra{\phi
+6\udot_{\mathsf{Sch}}}\ . \label{pmhSd}
\ea
The solutions may be readily found by making an educated guess.
Clearly, by spherical symmetry, the radiation has to move outward
radially and the expansion $\phi$ of the radial congruence $n^a$
therefore should be of the same form as for the Schwarzschild
spacetime but with a decreasing mass parameter. Since equation
\reff{pPSh} is identical with the analogous one for the
Schwarzschild spacetime, we are led to identify
$\udot_{\mathsf{Sch}}$ with the acceleration of a static
Schwarzschild observer, in particular, $\udot_{\mathsf{Sch}}$
represents the acceleration caused by the star's instantaneous
gravitational field. It is then straightforward to work out all
other quantities. The result is:
\ba
\phi &=&\fr2r\sqrt{1-\fr{2m(u)}{r}}\ ,\\
\theta &=& \fr{m'(u)}{r}\bra{1-\fr{2m(u)}{r}}^{-3/2} =
\sfr32\Sigma\ ,\label{thetasol}\\
\udot&=& \udot_{\mathsf{Sch}}-\theta; \qquad \udot_{\mathsf{Sch}} = \fr{m(u)}{r^2}\bra{1-\fr{2m(u)}{r}}^{-1/2}\ , \label{resacc}\\
\mu &=& -\fr{2m'(u)}{r^2}\bra{1-\fr{2m(u)}{r}}^{-1}= 3p = Q = \sfr32\Pi\ ,\label{resmu}\\
\E &=& -\fr{2m(u)}{r^3}\ .
\ea
It is worth pointing out that these solutions can  also be worked
out directly from the Vaidya metric \reff{metric} by choosing the
splitting unit vectors to be
\be
    u^a = \bra{1-\fr{2m(u)}{r}}^{-1/2}\bra{\fr{\partial}{\partial u}}^a
\ee
and
\be
    n^a = -\bra{1-\fr{2m(u)}{r}}^{-1/2}\bra{\fr{\partial}{\partial
    u}}^a + \bra{1-\fr{2m(u)}{r}}^{1/2}\bra{\fr{\partial}{\partial
    r}}^a,
\ee
respectively.

The obtained solutions may be interpreted as follows. First note
that from equation \reff{resmu} follows that a physically
acceptable, that is, positive energy density $\mu$ requires a
non-increasing (but otherwise unrestricted) mass, $m'(u)\leq 0$,
suggesting that a part of the star's mass is released as radiation
during collapse, with total luminosity $-m'(u)=4\pi r^2\mu$ at
infinity (in proper units). Further,  the expansion $\theta$ and
shear $\Sigma$ are negative since $m'(u)\leq 0$, meaning that the
spatial sections of the spacetime are contracting, which tends to
push nearby observers together. Equation~\reff{resacc} hence tells
us that an observer has to balance its always radially outward
directed acceleration according to the diminishing gravitational
attraction of the collapsing star and the growing contraction in
order to stay at rest. Finally, the expressions for sheet
expansion $\phi$ and tidal forces $\E$ are as for a star in
equilibrium but with a time-dependent mass due to mass-radiation
conversion during collapse.

\subsubsection{The perturbations}

With the background solutions for the Vaidya spacetime at hand, it
is now an easy task to rewrite  the covariant Regge-Wheeler
equations for the case of scalar and electromagnetic perturbations
into in terms of coordinates. From equation~\reff{RW2} one obtains
immediately for a massive scalar field $\psi$ with mass $M$,
$\psi=r^{-1}\Psi$, that \be \ddot\Psi -\hat{\hat \Psi}
-\udot\hat\Psi+\theta\dot\Psi -\bras{\E-M^2 + \delta^2} \Psi =0\ ,
\label{112rvad} \ee while equation \reff{Eregge} for the
electromagnetic perturbations, $\Es=r^{-2}E$, gives \be \ddot E
-\hat{\hat E} -\udot\hat E+\theta\dot E - \delta^2 E=0\ ,
\label{112remvad} \ee Introducing scalar spherical harmonics and
adopting the coordinates $\{r,u\}$ from above, we find the
corresponding Regge-Wheeler equation \be \fl
\bras{\bra{1-\fr{2m(u)}{r}}\fr{\partial^2}{\partial
r^2}-2\fr{\partial^2}{\partial u \partial r}
+\fr{2m(u)}{r^2}\fr{\partial}{\partial
r}+\bra{\fr{L}{r^2}+\lambda\fr{2m(u)}{r^3}+\lambda
M^2}}\mathscr{P}_{\Si}=0, \label{RWva} \ee where
$\mathscr{P}_{\Si}$ denotes the scalar harmonic component of
either $\Psi$ or $E$ and $L=\ell(\ell+1)$; the parameter $\lambda$
takes the value one for perturbations in a scalar field and the
value zero for electromagnetic perturbations.

It is instructive to compare the equations \reff{RWva} with the
corresponding equations \reff{RWscalar} and \reff{RWem} in the
Schwarzschild case. We found it most favourable to do the
comparison by employing coordinates $\{t,r_*\}$ which are defined
for $r>2m(u)$ by
\be
    t=u+r_*\ , \qquad r_*=r+2m(u)\bras{\ln\bra{\fr{r}{2m(u)}-1}}\
    .\label{newcoords}
\ee
In terms of the coordinates $\{t,r_*\}$, the equations \reff{RWva}
now become
\ba \fl \left\{
    \bras{-1-h(u,r)}\fr{\partial^2}{\partial t^2} +
    \bras{1-h(u,r)}\fr{\partial^2}{\partial r_*^2} -
    \fr{4m'(u)}{r-2m(u)} \bra{\fr{\partial}{\partial t}
    + \fr{\partial}{\partial
    r_*}}   -2h(u,r)\fr{\partial^2}{\partial t \partial r_*} \right.  \nonumber \\ \left.
    +
    \bra{1-\fr{2m(u)}{r}}\bras{\fr{L}{r^2}+\lambda\fr{2m(u)}{r^3}+\lambda
    M^2}\right\} \mathscr{P}_{\Si}=0\ , \label{newRWva}
\ea
where the function $h(u,r)$,
\be
    h(u,r)=4m'(u)\bras{\ln\bra{\fr{r}{2m(u)}-1}-\bra{1-\fr{2m(u)}{r}}^{-1}}\
    , \label{hur}
\ee
was introduced for brevity's sake. Clearly, when the mass $m(u)$
stays constant, the equations \reff{newRWva} reduce to the
familiar Regge-Wheeler equations \reff{RWscalar} and \reff{RWem},
respectively. Note that all new terms in the Regge-Wheeler
equations for the Vaidya spacetime \reff{newRWva} are proportional
to the mass change in time, $m'(u)$, and thus proportional to the
expansion $\theta$ [cf equation \reff{thetasol}]. This was to be
expected as, contrasting the Schwarzschild case, the only new
ingredient in the covariant Regge-Wheeler equations for the Vaidya
spacetime [see \reff{112rvad} and \reff{112remvad}] is
proportional to the expansion $\theta$ as well. Finally, the
accompanying Regge-Wheeler potentials retain the Schwarzschild
form but become time-dependent in the Vaidya case.

\subsection{More general spacetimes}

It was shown in \cite{Stewart-Ellis} that the metric of every LRS
class II spacetime can be given in diagonal form if local comoving
coordinates are chosen:
\be
  \fl  ds^2 = -A^{-2}(t,x)\,dt^2 + B^2(t,x)\,dx^2 +
    C^2(t,x)\,[\,dy^2+D^2(y,k)\,dz^2\,]\ , \label{LRSds}
\ee
where $D(y,k) = (\sin y\ , y\ , \sinh y)$ for $k = (1\ , 0\ , -1)$
labelling the closed, flat or open geometry of the sheet. The
kinematical quantities are now
\ba
    \phi &=& 2 \fr{\hat C}{C}\ ,\label{cphi}\\
    \udot &=& -\fr{\hat A}{A}\ , \label{cacc}\\
    \theta &=& \fr{\dot B}{B}+2\fr{\dot C}{C}\ ,\label{ctheta}\\
    \Sigma &=& \fr23\bra{\fr{\dot B}{B}-\fr{\dot C}{C}}\
    ,\label{csigma}
\ea
and equations \reff{Kdot} and \reff{Khat} can therefore be
integrated to yield the Gaussian curvature
\be
     K = \fr{C_1}{C^2(t,x)}\ ,
\ee
where the integration constant $C_1$ may be normalised such that
$C_1=k$. From equation \reff{Sigthetahat} or \reff{phidot} one
immediately obtains for the heatflux
\be
    Q = 2\bra{\fr{\hat{\dot C}}{C}-\fr{\dot B \hat C}{BC}}\ ,
\ee
while equation \reff{Raychaudhuri} directly yields the relation
\be
   \sfr12 \bra{\mu + 3p - 2\Lambda} = -\bras{\fr{\hat{\hat A}}{A} - 2\bra{\fr{\hat A}{A}}^2
    + \fr{\ddot B}{B} + 2 \fr{\ddot C}{C} + 2 \fr{\hat A\hat
    C}{AC}}\ , \label{mattercons}
\ee
which can be employed to constrain the metric functions such that,
e.g., $\mu +3p > 0$ holds. If a simple equation of state, $p =
\alpha\mu$, say, is assumed, one finds from this relation an
expression for the energy density $\mu$. On the other hand, the
field equations imply
\be
    \mu + \Lambda - K = \bra{\fr{\dot C}{C}}^2 -\bra{\fr{\hat
    C}{C}}^2 + 2\fr{\dot B \dot C}{BC} - 2\fr{\hat{\hat C}}{C}\,
\ee
which is in general different from the expression for $\mu$
obtained from equation \reff{mattercons} and therefore gives an
additional constraint for the metric functions. Finally, the
pressure variables and the electric Weyl curvature can be
calculated similarly but the resulting expressions are somewhat
long and will not be stated here.

\subsubsection{The perturbations}

When local comoving coordinates are chosen, the wave equations
governing the scalar and electromagnetic perturbations in the
sourcefree case are best given in form of equation \reff{112lrs}
and \reff{Eregge}, respectively, since there only the kinematical
variables enter and these are simple expressions in terms of the
metric functions [cf. equations \reff{cphi}--\reff{csigma}]. The
corresponding wave equations read in the case of scalar field
perturbations $\psi$ as
\be \ddot\psi -\hat{\hat
\psi} +\bra{\fr{\dot B}{B}+2\fr{\dot C}{C}}\dot\psi +
\bra{\fr{\hat A}{A}-2\fr{\hat C}{C}}\hat\psi + (M^2 -\delta^2)\,
\psi = 0\ , \label{scmetric} \ee while in the case of (rescaled)
electromagnetic perturbations $E$ (or $B$) they become
\be
 \ddot E -\hat{\hat E}  +\bra{\fr{\dot B}{B}}\dot E +\bra{\fr{\hat A}{A}} \hat{E}
 +\bras{3\fr{\dot B \dot C}{BC}-2\bra{\fr{\dot C}{C}}^2-\delta^2 } E
 =0.
 \label{Ereggemetric}\
 \ee
Once the metric functions $A$, $B$ and $C$ in \reff{LRSds} are
given for a known LRS class II solution, these wave equations may
readily be transformed into their concomitant coordinate
analogues.

Observe that the `radial' parameter $r$, covariantly defined by
the relations \reff{rdef}, agrees with the metric function
$C(t,x)$ in the line element \reff{LRSds} in account of
\reff{cphi} and \reff{ctheta}--\reff{csigma}. Thus the physical
electromagnetic perturbations $\Es$ are obtained from the rescaled
ones $E$ by the transformation $\Es=C^{-2}(t,x)E$. If the scalar
field $\psi$ is analogously rescaled as $\psi=C^{-1}(t,x)\Psi$,
then the wave equation for the scalar field perturbation
\reff{scmetric} takes the Regge-Wheeler form
\be \fl
 \ddot \Psi -\hat{\hat \Psi}  +\bra{\fr{\dot B}{B}}\dot \Psi +\bra{\fr{\hat A}{A}} \hat{\Psi}
 +\bras{M^2 + \fr{\hat{\hat C}}{C}-\fr{\dot{\dot C}}{C}  -\fr{\dot B \dot C}{BC}-\fr{\hat A\hat C}{AC}-\delta^2 }
 \Psi
 =0\ .
 \label{ScalarReggemetric}\
 \ee
It is thus obvious that once a harmonic decomposition has been
applied to the equations \reff{Ereggemetric} and
\reff{ScalarReggemetric}, the resulting wave equations will only
differ in their corresponding potential term.

If in addition to the LRS symmetry further symmetries  are
present, then the wave equations for the perturbations simplify
considerably. For example, for stationary LRS class II spacetimes
we have
\ba
 \ddot \Psi -\hat{\hat \Psi}   +\bra{\fr{\hat A}{A}} \hat{\Psi}
 +\bras{M^2 + \fr{\hat{\hat C}}{C}-\fr{\hat A\hat C}{AC}-\delta^2 }
 \Psi
 =0\ ,\\
\ddot E -\hat{\hat E}   +\bra{\fr{\hat A}{A}} \hat{E}
 -\delta^2  E
 =0\ ;
 \ea
while for spatially homogeneous spacetimes (where one may set
$A=1$) we have
\ba
 \ddot \Psi -\hat{\hat \Psi}  +\bra{\fr{\dot B}{B}}\dot \Psi
 +\bras{M^2 -\fr{\dot{\dot C}}{C}  -\fr{\dot B \dot C}{BC}-\delta^2 }
 \Psi
 =0\ ,\\
\ddot E -\hat{\hat E}  +\bra{\fr{\dot B}{B}}\dot E
 +\bras{3\fr{\dot B \dot C}{BC}-2\bra{\fr{\dot C}{C}}^2-\delta^2 } E
 =0\ .
 \ea

\subsubsection{Application -- LTB dust Universe}

 As an example of an inhomogeneous spacetime, let us consider the
 Lema\^{i}tre-Tolman-Bondi (LTB) \cite{Kramer-Stephani-MacCallum-Herlt,Lemaitre,Tolman,Bondi}
 dust Universe (with $\Lambda =0$), whose metric is given by
\be
       ds^2 = -dt^2 + \fr{(Y')^2}{k-\varepsilon f^2(r)}\,dr^2 +
    Y^2\, [\,d\vartheta^2+D^2(\vartheta,k)\,d\varphi^2\,]\ . \label{LTB}
\ee Here, a prime means $\partial/\partial r$ and a dot will
denote $\partial/\partial t$. Moreover, $Y=Y(t,r)$ is the solution
of $\dot Y^2 -2m(r)/Y = -\varepsilon f^2(r)$, wherein $\varepsilon
= (-1,0,1)$ corresponds to the hyperbolic, parabolic and elliptic
solution, respectively. The constraint equation can be integrated
completely, which yields an additional function $t_B(r)$.
Therefore, there are three functions which can be prescribed at
will: the `mass' $m(r)$, the `energy' $f(r)$ and the `bang time'
$t_B(r)$.

It follows from equations \reff{cphi}--\reff{csigma} that the
acceleration $\udot$ has to vanish and that the non-zero dynamical
variables take the form
\ba
    \phi &=& 2\ \fr{\sqrt{k-\varepsilon f^2}}{Y}, \\
    \theta &=&   \fr{\dot Y'}{Y'}+2\, \fr{\dot Y}{Y} \ ,\\
    \Sigma &=& \fr23 \bra{\fr{\dot Y'}{Y'}-\fr{\dot Y}{Y}}\ .
\ea
An expression for the energy density $\mu$ can be gained from
equation \reff{mattercons} and yields, using the above mentioned
constraint, \be \mu = \fr{2\,m'}{Y'\,Y^2}\ .\ee Finally, the wave
equations \reff{Ereggemetric} and \reff{ScalarReggemetric}
governing the perturbations $\mathscr P$ can be written as
\be \fl
    \ddot{\mathscr P} - \fr{k-\varepsilon f^2}{(Y')^2}\bras{\mathscr
    {P}'' - \bra{\fr{Y''}{Y'}+
    \fr{\varepsilon ff'}{k-\varepsilon f^2}} \mathscr{P}'}+\fr{\dot{Y}'}{Y'}\ \dot\mathscr{P} + (V-\delta^2) \mathscr P =0\
    ;
\ee
the potential $V$ is given by
\be
    V = V_{EM} = 3\ \fr{\dot Y\dot Y'}{YY'}-2 \bra{\fr{\dot
    Y}{Y}}^2
\ee
in the case of electromagnetic perturbations, whereas
\be
    V = V_{S} = M^2 +  \fr{m}{Y^3}+ \bra{\fr{\dot Y\dot Y' +\varepsilon ff'}{YY'}}
\ee
denotes the potential in the case of scalar field perturbations.

\subsubsection{Application -- Kantowski-Sachs dust Universe}

As a further example we consider a spherically symmetric
Kantowski-Sachs
 dust Universe with $\Lambda=0$ \cite{K,KS,Kramer-Stephani-MacCallum-Herlt}. The metric writes as
\be
    ds^2 = -dt^2 + B^2(t)\,dr^2 +
    C^2(t)\,d\Omega^2\ , \label{KS}
\ee
where the metric functions are conveniently expressed in terms of
a parameter $\eta(t)$ satisfying $dt=2Cd\eta$, namely
\ba
    B &=& m\bra{\eta\tan \eta +1} + b\tan\eta\ ,\\
    C &=& c\cos^2 \eta\ ; \qquad m\, , b\, , c = const \ .
\ea
Since this spacetime is spatially homogeneous, the acceleration
$\udot$ and expansion $\phi$ of the fundamental observer's
congruence have to vanish [cf. equations
\reff{cacc}--\reff{cphi}]. The remaining quantities are found to
be
\ba
    \mu &=& \fr{m}{BC^2} \ ,\\
    \cal E &=& \fr{m\cos^2\eta-3B}{3BC^2\cos^2\eta} \ ,\\
    \theta &=& c\ \fr{m\bra{\eta-3\sin\eta\cos\eta-4\,\eta\sin^2\eta}+
    b\bra{1-4\sin^2\eta}}{2BC^2}\ ,\\
    \Sigma &=& c\ \fr{m\bra{\eta+3\sin\eta\cos\eta+2\,\eta\sin^2\eta}+
    b\bra{1+2\sin^2\eta}}{3BC^2}\ .
\ea
The equations for the perturbations $\mathscr{P}$ now become in
terms of the parameter $\eta$
\be \fl
    \brac{\fr{\partial^2}{\partial \eta^2} -\fr{4\,C^2}{B^2}\ \fr{\partial^2}{\partial
    r^2}
    + \fr{B\bra{1+2\sin^2\eta}-m\cos^2\eta}{B\sin\eta\cos\eta}\
    \fr{\partial}{\partial \eta}
    +\ 4C^2\bra{V-\delta^2} } \mathscr{P}=0\ , \label{X}
\ee
where
\be
    V = V_S = M^2 + \fr{2B-m\cos^2\eta}{2BC^2\cos^2\eta}
\ee
in the case of scalar perturbations, and
\be
    V = V_{EM} =  \fr{3\,m\cos^2\eta-B\bra{3+4\sin^2\eta}}{2BC^2\cos^2\eta}
\ee
in the case of electromagnetic perturbations, respectively.

\section{Conclusion}

Employing the 1+1+2 formalism of Clarkson \& Barrett
\cite{Clarkson-Barrett}, we presented a covariant description of
LRS class II spacetimes in terms of scalar quantities, which all
have either a clear physical or geometrical meaning. We
investigated scalar and electromagnetic perturbations (test
fields) on LRS class II spacetimes and found that they are
governed by covariant wave equations [see equations \reff{RW2} and
\reff{Eregge}], which are the covariant generalisations of the
Regge-Wheeler equation, known to describe perturbations of the
Schwarzschild spacetime. In particular, it was shown that both
scalar and electromagnetic perturbations (in the absence of
sources) are governed by master equations of the same form, namely
the covariant \emph{generalised Regge-Wheeler equation},
\be
    \ddot \mathscr{P} -\hat{\hat \mathscr{P}} - {\cal A}\ \hat{\mathscr P} +\bra{\Sigma+\sfr13\theta}\dot \mathscr{P}
 +\bra{V-\delta^2}\mathscr{P} =0\ ,\label{conclusion}
\ee where the potential $V$ is different for the two cases
considered. To arrive at this specific form, one had to rescale
the perturbations with the `radial' parameter $r$, which is
induced from the Gaussian curvature $K$ of the sheet and can thus
only be defined covariantly for LRS class II spacetimes [cf
\reff{rdef}]. The findings have been discussed in some detail for
the particular cases of Schwarzschild and Vaidya spacetimes, and
for some dust Universe models. While the master equations may be
written simply as equation~(\ref{conclusion}) in covariant form,
they can become very untidy when written explicitly in
coordinates, demonstrating some of the advantages of using a
covariant approach.

\section*{Acknowledgements}

It is a pleasure to thank Mattias Marklund and Peter Dunsby for
their continuous advice and encouragement on this project, as well
as George Ellis and Charles Hellaby for discussions.

\section*{References}


\end{document}